\documentclass{article}

\usepackage{arxiv}

\usepackage[utf8]{inputenc} % allow utf-8 input
\usepackage[T1]{fontenc}    % use 8-bit T1 fonts
\usepackage{hyperref}       % hyperlinks
\usepackage{url}            % simple URL typesetting
\usepackage{booktabs}       % professional-quality tables
\usepackage{amsfonts}       % blackboard math symbols
\usepackage{nicefrac}       % compact symbols for 1/2, etc.
\usepackage{microtype}      % microtypography
\usepackage{graphicx}

\title{Vaccination strategies against COVID-19 and the diffusion of anti-vaccination views}

\author{
  Rafael Prieto Curiel\thanks{\url{https://www.peak-urban.org/people/rafael-prieto-curiel} and \url{https://rafaelprietoc.wordpress.com/}} \\
  Centre for Advanced Spatial Analysis\\
  University College London \\
  Gower Street, London, WC1E 6BT \\
  \texttt{rafael.prieto.13@ucl.ac.uk} \\
   \And
 Humberto González Ramírez \\
  Univ. Gustave Eiffel\\
  LICIT, F-69518\\
  Lyon, France\\
  \texttt{humberto.gonzalez-ramirez@univ-eiffel.fr} \\
}

\begin{document}
\maketitle

\begin{abstract}

Miss-information is usually adjusted to fit distinct narratives and can propagate rapidly through communities of interest, which work as echo chambers, cause reinforcement and foster confirmation bias. False beliefs, once adopted, are rarely corrected. Amidst the COVID-19 crisis, pandemic-deniers and people who oppose wearing face masks or quarantines have already been a substantial aspect of the development of the pandemic. With a potential vaccine for COVID-19, different anti-vaccine narratives will be created and, likely, adopted by large population groups, with critical consequences.

Here, we analyse epidemic spreading and optimal vaccination strategies, measured with the average years of life lost, in two network topologies (scale-free and small-world) assuming full adherence to vaccine administration. We consider the spread of anti-vaccine views in the network, using a similar diffusion model as the one used in epidemics, which are adopted based on a persuasiveness parameter of anti-vaccine views. 

Results show that even if an anti-vaccine narrative has a small persuasiveness, a large part of the population will be rapidly exposed to them. Assuming that all individuals are equally likely to adopt anti-vaccine views after being exposed, more central nodes in the network are more exposed and therefore are more likely to adopt them. Comparing years of life lost, anti-vaccine views could have a significant cost not only on those who share them, since the core social benefits of a limited vaccination strategy (reduction of susceptible hosts, network disruptions and slowing the spread of the disease) are substantially shortened.

\end{abstract}


\section{Introduction}

Major disruptions have been suffered at all social levels due to COVID-19. With over 5,000 daily confirmed casualties directly related to the virus, and the increased deaths from other causes including weakened healthcare systems \cite{owidcoronavirus} it will be among the top 10 causes of death of 2020. Surveys conducted in heavily affected cities suggest that still, a small part of the population has formed COVID-19 antibodies \cite{byambasuren2020estimating} and therefore, we are far from reaching herd immunity through natural infection \cite{fontanet2020covid}. Furthermore, with a conservative infection fatality ratio and an optimistic herd immunity threshold, several thousands of casualties would be suffered \cite{fontanet2020covid}. Although still not granted, chances are that a vaccine which grants some degree of immunisation will be available before herd immunity is reached.

Whilst a COVID-19 vaccine could be available at some point, its production and mass distribution would be a major challenge \footnote{For example, Robert Redfield, the Director of the US Centers for Disease Control and Prevention (CDC) mentioned on a US Senate hearing (September 16th, 2020) that he expects a vaccine available to the US public to return to ``regular life'' around the third quarter 2021 \url{https://cnb.cx/3c6n4rp}}. In most parts of the world, mass immunisation will not happen any time soon, and if a vaccine becomes available, targeted vaccination will be a key element to optimise the limited number of vaccines and mitigate the impacts of the pandemic. Under this scenario, still, unvaccinated individuals benefit from the immunity of vaccinated ones, as the reduction of susceptible hosts limits the spread of disease \cite{ma2013importance}. Likely, distinct vaccination strategies will gain prominence, particularly if the global demand for the vaccine is not matched with its production. Prioritise certain groups over others due to their exposure to the virus or their risk, will be a key policy, particularly during the early stages of the vaccine distribution.

Beyond the limitations related to the logistics of the vaccine, other social elements will become central in any strategy for coping with the pandemic. Already for COVID-19, a non-invasive element such as a face mask has been highly rejected by some, whilst protests and demonstrations around the world against responses to the pandemic are frequent. Anti-lockdown protesters and conspiracy theorist have already demonstrated against the possibility of a mandatory vaccine \footnote{For example, a protest in London in August 29th, 2020 included coronavirus sceptics, 5G conspiracy theorists and so-called ``anti-vaxxers'' \url{https://bit.ly/3kvsyiJ}}. A survey conducted in 27 countries indicates that 1 in 4 people would not get a COVID-19 vaccine \footnote{For reasons such as worry about side effects and doubts about its effectiveness \url{https://bit.ly/3mSEXPA}} and another survey suggests that around half of the population would rather let others receive the vaccine first \footnote{From a survey in Mexico, available at \url{https://bit.ly/334xxkr} which could indicate both fears and solidarity with others}. Misguided safety concerns have already reduced vaccination coverage, causing the re-emergence of some previously controlled diseases \cite{andre2008vaccination}.
Opposition to vaccination could amplify outbreaks \cite{larson2020lack} as happened for measles in 2019 \cite{johnson2020online}. Some people fear more a mandatory vaccine for COVID-19 than the virus itself \footnote{See \url{https://www.texasmonthly.com/news/texas-anti-vaxxers-fear-mandatory-coronavirus-vaccines/}}. Anti-vaccination views offer a wide range of narratives (such as safety concerns, conspiracy theories and the use of alternative medicine) and it has been predicted that these views could dominate in a decade \cite{johnson2020online}. Different narratives against the vaccine would friction vaccination strategies, and the adherence pattern can greatly influence the coverage and time needed to eliminate transmission \cite{hardwick2020dynamics}.

Either because of logistic restrictions and/or because people refuse to be vaccinated, strategies in which some people -perhaps a small percentage- are immunised, need to be considered in order to reduce the impacts of the pandemic. Here, we consider a network-based population model and five vaccination strategies with different rates of vaccination. A modified version of the SIR model, SVIR (due to the existence of the Vaccine) is used to detect the long-term behaviour of a pandemic, where the efficiency of each vaccination strategy is measured using the years of life lost and the time to eliminate transmission. Simulating the evolution of the pandemic under different network topologies allows us to compare vaccination strategies, adherence patterns and detect successful immunisation strategies. Using a similar technique, we consider the dispersion of anti-vaccination views on a network, which pass through individuals with a certain level of \emph{persuasiveness}. Assuming that all individuals are equally likely to adopt anti-vaccination views after being exposed to them, we demonstrate that central individuals are much more likely to adopt these views. Considering that individuals which share anti-vaccination views refuse a vaccine, we quantify the costly impact it has on optimal vaccination strategies. 

\section{Modelling vaccination and anti-vaccination views on a network} %%% lit

\subsection{Dealing with non-universal vaccination}

Vaccinations changed dramatically our health and quality of life. Yearly cases of measles, mumps or smallpox decreased by more than 98\% since the introduction of a vaccine \cite{ma2013importance}. Formerly fearsome diseases are now rare in many parts of the world thanks to vaccination programs \cite{andre2008vaccination}. Yet, due to a limited number of vaccines, their cost, views against vaccines and many more reasons (including the fact that with more vaccines, the spreading of a virus slows down, which in turn, tends to discourage more people to participate in the vaccination campaign \cite{cai2014effect}) universal vaccination is almost impossible.

Understanding the effects of limited vaccination, thus, is vital and likely will play a relevant role if a vaccine against COVID-19 becomes available. Vaccination strategies have two core objectives: lower and delay the peak size (``flattening the curve'') and reduce the final infected population to limit morbidity \cite{feng2011modeling}. There are, however, many challenges for designing a vaccination strategy. Deaths can be prevented by first targeting highly vulnerable populations \cite{fontanet2020covid} (such as elderly people, with severe morbidity, or vulnerable communities, including prison and homeless populations \cite{randolph2020herd}). Yet, targeting vulnerable population might not reduce much the viral circulation and might not reduce the final infected population either, particularly if the targeted population has limited social contacts. Another strategy is to target highly exposed populations, such as health-care workers or with frequent social contacts, as it should slow down the overall exposure to the virus. Other strategies, for instance, targeting young people (as more years of life are saved with each vaccine), random people (as means of some distributive fairness) could also be proposed for a variety of reasons. Thus, the relevance of being able to compare the expected outcomes of different vaccination strategies.

\subsection{Modelling immunisation in a population}

One of the first mathematical models in epidemiology was concerned with immunisation. In 1760, Daniel Bernoulli predicted the impact of immunisation with cowpox upon the expectation of life of the immunised population \cite{scherer2002mathematical}. Since then, many models of immunisation have been constructed. Particularly, the Susceptible, Infected and Recovered (SIR) compartments model proposed in 1927 by William Ogilvy Kermack and Anderson Gray McKendrick \cite{husein2020modelling} has been extended for the application of a vaccine (SVIR) \cite{liu2008svir, feng2011modeling, el2019analysis, ehrhardt2019sir, luong2019mathematical}. 

Nearly 100 years ago, the lack of computational power meant that dealing with a high number of variables was nearly impossible and obtaining numerical results was very costly and so compartmental models in epidemiology were translated, under many assumptions, into differential equations. Algebraic solutions were more accessible than numerical results, and gave rudimentary insights of the general behaviour of the dynamics of the disease \cite{chauhan2014stability}. For instance, this basic model can give us a rough idea of the number of individuals that would need to be infected to achieve herd immunity \cite{randolph2020herd}. In the case of a vaccine, it was showed that uniform vaccination is always less effective than targeted vaccinations and the optimal strategy involves vaccinating certain individuals first \cite{anderson1992infectious}.

Many of the implications of the SIR and the SVIR models rely on simplifying assumptions, such as an homogeneous population mixing and uniform recovery rate. In recent years, thanks to the gain of computational power and the development of specific tools and frameworks, some of these assumptions have been relaxed, particularly assuming an homogeneous population with homogeneous contacts. The use of networks in epidemiology is a powerful tool to relax both as tagged nodes allows considering disease dynamics for heterogeneous populations \cite{nasution2020model}, and distinct distribution of the edges allows considering distinct types of contacts. For example, in the case of a sexually transmitted infection, the sexual partnership network is a natural framework for modelling the disease \cite{lloyd2007network}. The contact network among people is a strongly connected small-world-like graph \cite{SmallWorldStrogatz, BarabasiStructureNetworks}(meaning that path in the network needed to connect any two nodes is rather small, or with many short-cut connections \cite{mosbah2019dynamics}) with a well-defined scale for the degree distribution of the number of contacts of each node \cite{eubank2004modelling}. The network structure has a major impact on the spread of infectious diseases and therefore, on successful vaccination strategies \cite{hartvigsen2007network, ma2013importance, lloyd2007network, zhao2007simulating}. The basis of the disease models on a network is a compartments model (as with the SIR and SVIR model), but minor changes on the network's connectivity might alter results significantly. Furthermore, the network structure appropriate for a given setting not only depends on the structure of the contacts of the population (which likely change over time), but on the infection itself \cite{lloyd2007network}.

Designing successful immunisation strategies need to take into account the inhomogeneous connectivity properties of the network \cite{pastor2002immunization}. An infected person, even with a reduced number of contacts, can pass the virus between separate clusters, particularly if any of such contacts is a short-cut connection in the network \cite{eubank2004modelling}. On a small-world network, all individuals, even if all of them have a small number of contacts, are within a few infections to be infected themselves. On a different network structure, some nodes (often called ``hubs'') have many more connections than others and the network as a whole has a power-law degree distribution, as in the case of the network of sexual contacts, which exhibits scale-free features \cite{pastor2002immunization}, referred to as a scale-free network \cite{BarabasiEmergence}. On a scale-free network, uniform vaccination is always less effective than targeted vaccination \cite{lloyd2007network}. Scale-free networks are resilient to a few disconnections but are strongly affected by selective node damage. If a few of the most connected nodes are removed, the infection suffers a substantial reduction in its ability to propagate \cite{pastor2002immunization}. Control programs should be targeted towards the highly connected nodes, and such programs will be much more effective than those that target nodes at random \cite{may2001infection}.

These strategies, however, assume that all individuals would accept the vaccine if offered. A targeted vaccination towards the most connected nodes will be highly ineffective if, for any reason, some of the central nodes refuse a vaccine. The impact of those who reject a treatment has been analysed among distinct demographic groups \cite{hardwick2020dynamics} which showed that non-adherence to a treatment, such as a vaccine, can greatly influence the needed coverage to eliminate transmission. In the case of network structures, the impact of non-adherence has not been explored and it is likely to play a significant role if a vaccine against COVID-19 becomes available.

\subsection{Opinion dynamics and anti-vaccination views}

An idea or an opinion, such as views in favour or against a vaccine, is transmitted -frequently as an intentional act- from one person to another \cite{BettencourtGoodIdea}. Through distinct types of interactions, people want to persuade others to adopt an opinion \cite{FSmithDynamicsPersuasion}. The views of others might have an impact on individual beliefs \cite{LataneSocialImpact} who update their own opinion. Different ways to model, not only different perceptions or ideas, but to also the updating process of those ideas -obtained through interactions with others or with some externalities- aim at capturing why opposing views (which could be the acceptance of a vaccine), can emerge and co-exist in a society, even if all individuals try to reach a consensus.

Opinion dynamics has been modelled through a variety of angles and techniques, for instance, kinetic models of opinion formation \cite{GalamDynamics1982}, mean-field analysis, which usually leads to a system of differential equations \cite{BoltzmannLeaders}, agent-based models and even epidemiological models \cite{BettencourtGoodIdea}. Frequently, two opposing opinions are assigned to the extremes of an interval, say, $[-1,1]$, individual opinions are modelled as a number $s$ in that interval, according to the position with respect to the two opinions, and through interactions, individuals have some compromise (opinions get closer) and other elements, such as memory loss \cite{ModellingFearRPC}, the presence of leaders \cite{BoltzmannLeaders}, the ability to convince others with a different opinion \cite{StatisticalPhysicsSocialDynamics}, varying levels of assertiveness \cite{BoltzmannSegregation}, the fact that more extreme opinions are more difficult to change \cite{ToscaniBinary}. Also, confrontation with distinct views and opinions does not happen at random. Either because of the dynamic process through which opinions are updated \cite{StatisticalPhysicsSocialDynamics}, or because of preferential interactions, people tend to be surrounded by others with similar views \cite{BoltzmannSegregation}.

As in the case of the diffusion of a virus, different network structures are a natural framework to analyse opinion dynamics. Ideas an opinions tend to propagate on a network between adjacent nodes \cite{DeffuantModel}. For instance, opinion leaders, who have a greater impact, can be the hubs of the social (scale-free) network \cite{BoltzmannLeaders}, whereas the distribution of online contents can pass through highly connected nodes on a (small-world) network \cite{FakeNewsOnline}. The results observed of complex opinion dynamics are that a population might have relevant levels of polarisation or fragmentation \cite{OpinionSimulations}, people tend to have interactions with others with similar views, and to they are more frequently exposed to confirmation bias \cite{FakeNewsOnline}. People tend to select and share content according to a specific narrative and to ignore the rest \cite{FakeNewsOnline}. Users tend to aggregate in communities of interest, which causes reinforcement and fosters confirmation bias, segregation, and polarisation and leads to the proliferation of biased narratives fomented by unsubstantiated rumours, mistrust, and paranoia \cite{FakeNewsOnline}.

Massive misinformation is becoming one of the main threats to our society \cite{FakeNewsOnline, larson2020lack}, and if a vaccine for COVID-19 becomes available, chances are that opposing views towards immunisation will become the crucial element, especially if the propagation of the virus itself and if vaccination strategies are, in any way, entangled within the network of individuals who refuse it.

\subsection{Anti-vaccination views} 

%%% ATIVAX

Opposition to vaccination has amplified outbreaks, as happened for measles in 2019 \cite{johnson2020online}. One-third of {US} parents plan to skip flu shots for their kids \cite{thomas2018one} and thus, kids in these {US} hot spots at higher risk because parents opt out of vaccinations \cite{sun2018kids}. Vaccine safety concerns receive more public attention than vaccination effectiveness \cite{andre2008vaccination}. According to a study published in August 2020, nearly one in four adults would not get a vaccine for COVID-19 \cite{boyon} and in some countries, more than half of the population would not get it, including Poland and France.

%More on AV views - WIP.

%%% vaccine for covid
Currently, a vaccine for COVID-19 is not granted and there are important questions. Still, we do not know whether the immunisation of a vaccine would wane over time and so how long the acquired immunity would last \cite{randolph2020herd}. Even with a vaccine, if reinfection could occur, persistent herd immunity may never be attained, which then could lead to cyclic outbreaks \cite{kissler2020projecting, randolph2020herd}. Finally, we do not know if a sufficiently large population would reject the vaccine delaying even more the process of obtaining herd immunity. 

\section{Methods}

A population of $N$ individuals is constructed with two attributes: current age (or simply age) and age at death (or simply death). Both age and death are sampled from a scaled Beta distribution $100Beta(\alpha_a, \beta_a)$ and $100Beta(\alpha_d, \beta_d)$ for age \cite{SILVA2020110088} and death respectively, corresponding to their age today and their age at the moment of their death. Individuals with death smaller than age are resampled. After resampling, with $\alpha_a =2$ and $\beta_a= 3$ gives a median age of nearly 35 years and with $\alpha_d =5$ and $\beta_d= 2$ gives a median age at the moment of death of 75 years (close to the median age and expected life in Mexico City).

Two network topologies are considered, where the nodes represent the individuals and the edges their contacts. Firstly, a small-world network with a rewiring probability of $r$ and secondly, a scale-free network with power $p$. Although very young or elderly population tend to have fewer contacts, we construct age and network independently, for simplicity.

A SVIR model is constructed in the network as follows. Firstly, some randomly-selected nodes are infected and the rest of the nodes are susceptible. After the initial infection process, a percentage $\nu$ of the individuals is vaccinated, called the \emph{vaccination rate}. If the individual was susceptible, she or he gains permanent immunity and is no longer capable of passing the virus, but if the person was already infected, the vaccine does nothing (a scenario which could happen for individuals who are asymptomatic and receive a vaccine). Each time step, which could be considered a day, susceptible individuals who are adjacent to an infected node are also infected, with a probability $\pi$. After $\tau$ steps in which the node has been infected, the node is moved to the recovery state and no longer passes the virus to others. The process stops when the population has no infected nodes, and that time is noted as $T$.

After infection, some of the individuals who ``recover'' might not survive. It has been noted that, in the case of COVID-19, lethality increases with age. For simplicity, we consider a linear impact of age, so that a person with $y$ years does not survive with a probability of $\phi y$, for some $\phi>0$. The average number of years of life lost due to the pandemic is also noted, as $D$, computed as the sum of the number of years that people who passed away lost (that is, their death minus their current age), divided by the population size $N$.

\subsection{Vaccination strategies}

Five target-vaccination strategies, based on the age or on the network attributes are considered, for a portion $\nu$ of the individuals who receive the vaccine:
\begin{itemize}
\item \emph{Degree} - target the top $\nu$ nodes with a higher degree. More connected nodes receive the vaccine first. 
\item \emph{Centrality} - vaccinate the top $\nu$ nodes with the highest node betweenness.  
\item \emph{Peripheral} - apply the vaccine to nodes with the lowest node betweenness.
\item \emph{Age} - vaccinate the top $\nu$ nodes with highest age (elderly population) first.
\item \emph{Random} - select a portion $\nu$ of the individuals at random.
\end{itemize}

The first three strategies (Degree, Centrality and Peripheral) are based on network attributes, meaning that nodes are sorted based on their degree or the node betweenness and the top $\nu$ nodes are vaccinated. The Age strategy is based on node attributes (their age) and the last strategy (Random) uses no information about the node.

\subsection{Comparing vaccination strategies}

We consider two distinct metrics for comparing vaccination strategies. Firstly, from the population of $N$ individuals from which a small group is initially infected, vaccines are applied according to some strategy, and the SVIR dynamic is simulated until no individuals are infected. From the recovered individuals, the casualties are simulated and the average number of years of life lost $D$ is computed.

A second metric to understand the impact of distinct vaccination strategies is the time needed for the dynamic to stop (when the population has no infected nodes), $T$.

Both $D$ and $T$ are computed for a vaccination strategy and for a certain vaccination rate $\nu$, so that $D_{Age}(\nu)$ and $T_{Age}(\nu)$ are reported for the Age strategy with a vaccination rate of $\nu$, and likewise for other strategies.

For a given network and for a fixed vaccination rate and a (non-random) strategy, results might vary for two reasons. Firstly, the initial population that was infected with the virus is sampled from the population. But secondly, because of the transmission of the virus itself. Susceptible individuals who are adjacent to an infected node are infected with a probability $\pi$ on each time step, and so the progression of the virus might change between different realisations, even with the same initial conditions. Thus, we simulate each vaccination strategy 500 times for different vaccination rates $\nu$ and report the intervals which contain $D_{S}(\nu)$ and $T_{S}(\nu)$ for different strategies $S$. 

Here, we consider a perfect vaccine, which grants permanent immunity, although it is possible that the level of antibodies of an immune person drops after a certain time below a critical threshold and so there could be waning immunity \cite{ehrhardt2019sir, scherer2002mathematical}, which could take individuals back into being susceptible. 

\subsection{Diffusion of anti-vaccine views}

The diffusion process of anti-vaccine views plays a relevant role in the vaccination process and therefore, in the evolution and the burden of the pandemic. The social network is a crucial part of the diffusion process, as anti-vaccine views are spread through contacts, convincing other individuals of their own views of the vaccine.

We construct a diffusion process on a network, similar to the SIR model, but for the adoption of an idea \cite{BettencourtGoodIdea} as follows. Initially, all individuals are susceptible to the views and some randomly selected individuals have anti-vaccine views. Each time step, individuals with anti-vaccine views share them with their contacts. Individuals who are exposed for the first time to anti-vaccine views make a permanent decision based on the persuasiveness $\theta$ of the ideas, where $\theta \in [0,1]$ are the chances that the individual is persuaded by anti-vaccine views. A small value of $\theta$ means that most individuals are not convinced by anti-vaccine views, and the opposite for a high value of $\theta$. Since we assume that the decision of people is permanent, then on the first step, only the neighbours of those who initially share anti-vaccine views are exposed to them. On the second step, only the neighbours of individuals who were convinced on the previous step are then exposed to anti-vaccine views, and so on. Each time step, only the neighbours of individuals who were convinced on the previous step are exposed for the first time to anti-vaccine views and make a decision. The dynamic stops when no new individuals are convinced of anti-vaccine views.

At the final stage of this dynamic, some individuals have anti-vaccine views, some have been exposed to anti-vaccine views but do not support them, and some individuals (potentially a large group or none, depending on the persuasiveness of $\theta$) were never exposed to the anti-vaccine views. We assume that only individuals who share anti-vaccine views reject a vaccine, and the rest, whether they were exposed or not to the views, could be immunised. 

\section{Results}

\begin{figure}[h]
  \centering
\includegraphics[width=0.95\textwidth]{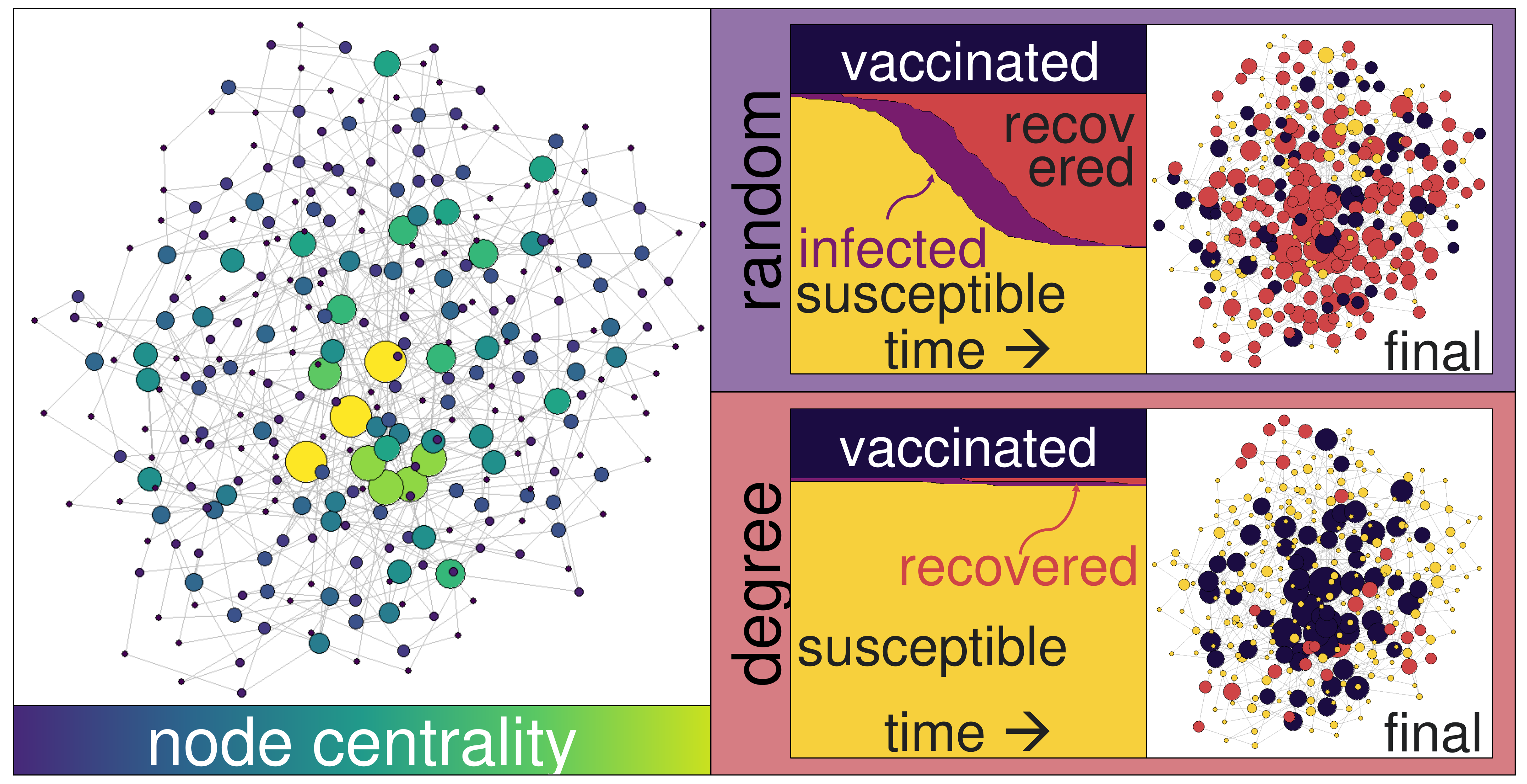}
  \caption{Schematic representation of a scale-free network with power $p=1$ (left) which has larger and brighter nodes displaying their node betweenness. The evolution of the SVIR model for a Random vaccination strategy and the Degree vaccination strategy, with the same vaccination rate $\nu = 20\%$ (centre) shows that, after time, the majority of the population remain susceptible with a targeted Degree vaccination strategy, but is recovered with a Random strategy. The final state of the nodes (right) shows that many nodes are recovered (red) with a Random strategy, but most remain susceptible (yellow) with a Degree strategy. }
  \label{Network}
\end{figure}

Considering a scale-free network with $N = 20,000$ individuals, with 250 infected individuals at time $t = 0$, targeted vaccinations yield drastically distinct results than a random vaccination strategy (Figure \ref{Network}). Applying the vaccine to central nodes not only protects them from receiving the virus, but also, slows down the diffusion process of the virus, allowing nearby infected nodes to recover and stop passing the infection to others (as it has been observed previously \cite{pastor2002immunization, may2001infection}). The final size of the recovered population varies considerably. For example, with a random vaccination strategy and with $\nu = 20\%$, a large part of the population would be infected at a point. With targeted vaccinations, the size of the infected population drops drastically and so the majority of the individuals remain susceptible, even with the same (small) vaccination rate $\nu = 20\%$.

We compute the (simulated) years of life lost and observe that without a vaccine, the population expects to suffer up to 4.7 years of life lost due to the pandemic, and with a universal vaccine, the loss is negligible. Yet, for a rate of $\nu = 60\%$ of randomly vaccinated people has the same impact, in terms of life saved than $\nu = 30\%$ if targeted correctly with a Degree or Centrality strategy.

\begin{figure}[h]
  \centering
\includegraphics[width=0.95\textwidth]{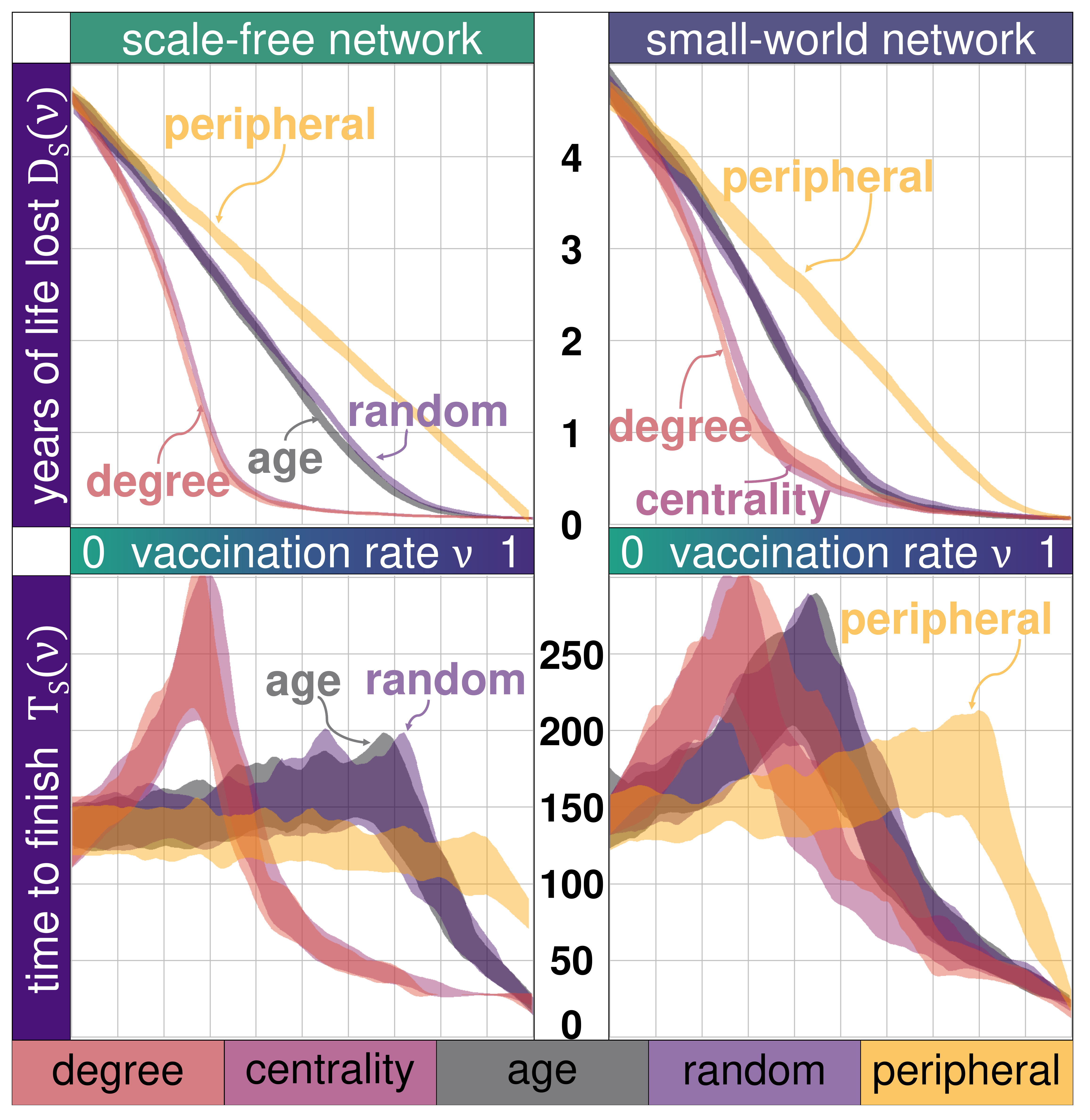}
  \caption{Years of life lost $D_{S}(\nu)$ (vertical axis) according to the vaccination rate $\nu$ (horizontal axis) with different vaccination rates (in the top panel) and the corresponding time $T_{S}(\nu)$ (vertical axis) with the same vaccination rate (in the bottom panel). Simulations in the left are on a scale-free network (with power $p=1$) and simulations in the right are on a small-world network (where the rewiring probability is $r=0.1$). Each strategy is simulated 500 times with $N = 20,000$ individuals, with a probability of infecting neighbouring nodes at each step of $\pi =0.05$ and with a varying vaccination rate $\nu$. The lethality of the virus increases with $\phi = 0.05$, meaning that a person with $y$ years does not survive after being infected with a probability of $0.05 y$.}
  \label{Metrics}
\end{figure}

In a population in which some individuals are much more connected than others (a scale-free network) or one in which individuals are just a few nodes away from each other (a small-world network), targeting nodes with a high number of neighbours (Degree strategy) or nodes with a high betweenness (Centrality strategy) yields similar results in terms of the years of life saved (Figure \ref{Metrics} top panel). 

Surprisingly, targeting the most vulnerable population (in our model, elderly population, which are more likely to pass away after the infection) does not give better results in terms of the years of life lost than a random strategy, as it does not slow down the diffusion process of the virus as much as the Degree of the Centrality strategy. The Peripheral strategy gives the worst results in terms of the years of life saved, as it targets the nodes less capable of slowing the diffusion of the virus. The Age strategy does perform better than a Random strategy in terms of the number of casualties, but not in terms of the years of life lost, which highlights that for vaccination strategies, measuring casualties only does not give sufficient information and the demographics (age) of the population who passed away should be taken into account. The Age strategy targets individuals who are more likely to pass away, but it does not save much more years of life than the Random strategy, as it prevents casualties from people with less expected life and this is ignoring potential correlations between age and centrality. However, elderly people tend to have fewer contacts \cite{fontanet2020covid} and so targeting them with a small vaccination rate could imply some peripheral vaccination as well.

Vaccination, particularly with a small rate $\nu$, slows down the evolution of the virus, so it also slows down the time for the process to finish (Figure \ref{Metrics}, bottom panel). Without vaccination, it takes around 130 steps for the virus to spread across the population and for them to recover. Applying a vaccine to the most connected or the most central nodes (Degree and Central strategy) prevents the virus to spread to many nodes, but still, it moves through the network between less central nodes, taking up to twice as many steps as compared to the case with no vaccine. The Age and the Random strategy also friction the evolution of the virus and in turn, it also takes longer for the pandemic to end. With Degree, Centrality, Age and even the Random strategy, applying the vaccine to only some individuals (more than half of the individuals in some cases) the vaccine delays the end of the pandemic. Only if a very large part of the population gets vaccinated, the pandemic finishes faster than without any vaccine.

The Peripheral strategy on a scale-free network accelerates the end of the pandemic (with all vaccination rates) since it immunises individuals which would be infected last, as opposed to the case of a small-world network. On a small-world network, less connected nodes still pass the virus to its neighbours so small vaccination rates delay that process and thus, increase the time $T_{S}(\nu)$.

Notice that although the SVIR model is a dynamic process, it is based on a static network of individuals, meaning that no new connections are formed. However, if the vaccine process takes long, it is likely that new contacts (more edges) are added to the network, with its possible implications in terms of the years of life lost $D_{S}(\nu)$ and the time  $T_{S}(\nu)$.

\subsection{The topology of the network}

The topology of the social network plays a relevant role in detecting the optimal vaccination strategy and its impacts in terms of saving lives and speeding the process of the pandemic. Targeting the most central or the most connected nodes on a scale-free network rapidly decrease the years of life lost with some vaccination rate, but it is slightly less effective on a small-world network.

Qualitatively speaking, except for the Peripheral vaccination strategy, both networks show similar results. Targeting the vaccine to the most central or the most connected nodes reduces its casualties, although for small vaccination rates, increases the time for the pandemic to finish. Targeting the most vulnerable individuals (even assuming that they are as central or connected as other individuals, which might not be true) has an impact as reduced as a Random vaccination strategy and still, for small vaccination rates, both will increase the time $T_{S}(\nu)$. When central nodes are removed through vaccination, fewer individuals will be infected, but it also takes longer to propagate among those individuals who get the virus, so the time  $T_{S}(\nu)$ nearly doubles as compared to the scenario with no vaccination. 

A strategy targeting the younger nodes could also be designed, but due to their small lethality, that strategy saves fewer years of life and prevents less casualties than any other strategy, and also, the time $T_{S}(\nu)$ of such strategy follows the same patterns than the Age and the Random strategies since none of those strategies use the properties of the network.

\subsection{Anti-vaccine views}

\begin{figure}[h]
  \centering
\includegraphics[width=0.7\textwidth]{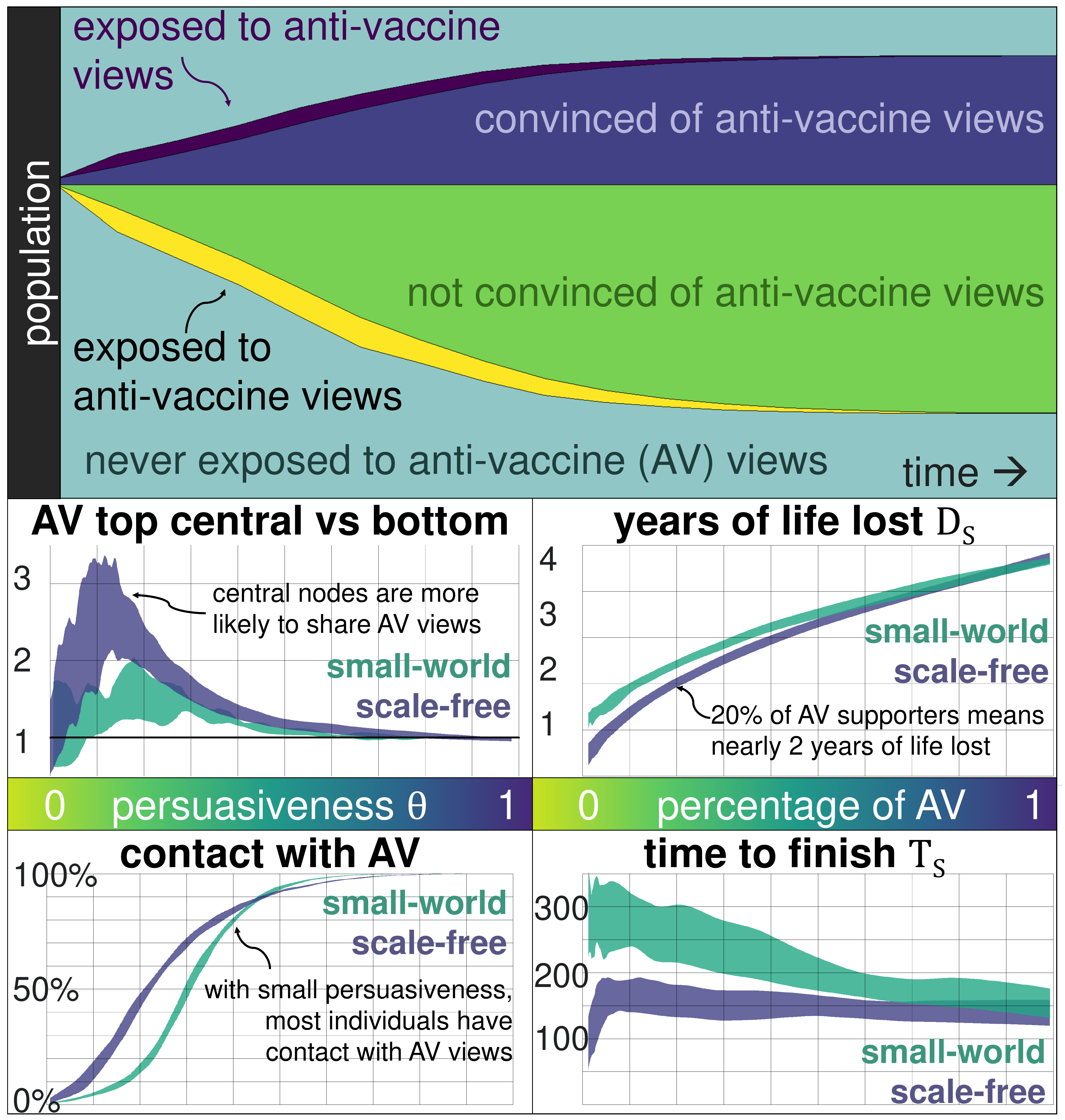}
  \caption{Anti-vaccine views are shared on a network, with $N=20,000$ individuals and where some randomly-selected nodes initially share the views. The top panel shows the evolution of those who are exposed for the first time to anti-vaccine views (dark purple) and are convinced by them (dark blue), those who are exposed for the first time (yellow) but do not adopt them (green) and individuals who never have contact with them (light blue). The odds of a person sharing anti-vaccine views, comparing the top 10\% most central nodes against the bottom 10\% (central left panel) shows that for extreme values of $\theta$, centrality does not have an impact (since all individuals share similar views), but for medium values, the most central nodes are two or more times more likely to share anti-vaccination views since they are more exposed. With values of $\theta$ between 0.1 and 0.3, more central nodes are more likely to be exposed to the anti-vaccination views and, since individuals are equally likely to adopt the views after their first exposure, more central individuals are more likely to adopt them. The impact is more pronounced on scale-free network (blue) than on a small-world network (green). By the end of the diffusion process of the anti-vaccination views (bottom left panel) even with a small persuasiveness $\theta$, most individuals will have contact with those views. On a scale-free network, with a persuasiveness of $\theta = 0.3$, around 70\% of the population has contact with the views (even if the majority of them rejects them and do not pass them onwards). The impact of the size of the anti-vaccine community is relevant in terms of the years of life lost $D_{S}(\nu)$ (central right)  and has some impact in terms of the time to finish $T_{S}(\nu)$, considering a vaccination rate of 30\%. Assuming a Degree vaccination strategy, we see that if 20\% of the people support anti-vaccination views, nearly two years of life are lost (so that $D_{S}(0.3) \approx 2$ years), although with a vaccination rate of 0.3 and the Degree vaccination strategy, only 0.3 years of life are lost. }
  \label{Antiva}
\end{figure}

Anti-vaccine views strongly depend on the persuasiveness $\theta$. For small values of $\theta$ the idea dies fast and only few individuals ever share those views (Figure \ref{Antiva}). As with an infection, most of the individuals remain ``suceptible''. For medium values of $\theta$, a large part of the individuals will have heard of anti-vaccination views, although many of them will not be convinced by them. Only with high values of $\theta$, anti-vaccination views percolate the network. The final size of the anti-vaccination community depends in a non-linear way on the persuasiveness $\theta$. 

More central nodes are more exposed to anti-vaccination views than peripheral nodes (Figure \ref{Antiva}), particularly if the network has hubs (scale-free). In turn, assuming that nodes are equally likely to adopt anti-vaccination views the first time they are exposed to them, more central nodes are more likely to share anti-vaccination views. The top 10\% most central nodes are between two and three times more likely to share anti-vaccination views than the 10\% least central nodes for small (but still larger than zero) persuasiveness $\theta$.

For small values of persuasiveness $\theta$, most individuals have contact with the anti-vaccination views, even if they reject them and do not pass them onwards (bottom left panel of Figure \ref{Antiva}). This can be particularly challenging, as anti-vaccine views have many narratives (including different conspiracy theories, safety concerns, the use of alternative medicine, medical risks such as autism \footnote{See, for instance, \url{https://www.healthline.com/health/vaccinations/opposition}}). Each narrative might follow similar dynamics, reaching most individuals but convincing only some, triggering collective narratives and creating echo chambers which reinforce themselves. Thus, the diffusion of anti-vaccination views could also be conceived with many realisations of the same dynamic, each one for separate narratives which convince distinct susceptible individuals.

The final size of the anti-vaccine community is directly related to the persuasiveness parameter $\theta$ and the number of steps needed for the dynamic to stop is surprisingly short. With a small persuasiveness $\theta$, the idea dies fast; with a large $\theta$, the idea percolates in just a few steps. Only with medium values of $\theta$ there is some delay, but still, is fast (around 10\% of the steps needed for the virus dynamics). A virus might propagate fast, but fake news and miss-information propagate many times faster.

Finally, assuming that the network in which opinions are shared is the same network in which the virus propagates (which is a strong assumption, considering that opinions can be shared online or other ways which does not require physical contact), the impact of the anti-vaccination views is highly relevant. Considering a vaccination rate of 30\% and a Degree vaccination strategy, 0.3 years of life are lost on a scale-free network and little more than a year on a small-world network, but that could increase to nearly two years if only 20\% of the population shares anti-vaccination views (Figure \ref{Antiva}, right panels). Individuals with a high centrality are the key element of the optimal vaccination strategies, as they slow down the diffusion process of the virus, but their centrality also places them as nodes more likely to share anti-vaccination views, thus, having a substantial cost on the vaccination strategy. A larger anti-vaccination community reduces the time to finish $T_S$ as the virus propagates faster, except for a very small size, on a scale-free network only.

\section{Discussion}

A vaccine for COVID-19 will not be a silver bullet to end the pandemic and mitigate its impact (if it exists and is safe). Beyond the logistics related to producing and distributing billions of vaccines around the world, chances are that most countries would face limited availability of the vaccine, especially at early stages. Still, outbreaks can be contained by a strategy of targeted vaccination combined with early detection \cite{eubank2004modelling}.

Our simulations cannot be used to estimate the years of life lost we would experience due to COVID-19 for different vaccination rates since that requires considering comorbidities, access to health services and more factors, but with a simulated population and a pandemic, we obtain a qualitative description of how the process might evolve.

The main benefit of a person being vaccinated does not rest just on the immunity that they gain, but also on the fact that they stop spreading the virus to others. As such, targeting vaccination to more exposed people, have more contacts and are more likely to pass the virus onwards is the strategy with the highest social gain. At early stages of vaccination, healthcare staff should be the priority, and also, other professions with a high number of contacts and which have a high centrality on the network, such as a person who works at a shop, a barber, a security guard or a taxi driver, as they have frequent contact with many different clusters of people.

Measuring the burden of the pandemic in terms of the estimated years of life lost and not just on the casualties allows us to put in perspective the impact of targeting elderly population first. Although it could reduce the number of casualties, it does not reduce the years of life lost compared to a Random strategy of vaccination.

\subsection{Anti-vaccine views will be highly relevant}

If a vaccine for COVID-19 becomes available, anti-vaccine views and in general, fake news related to the virus, will then be a second pandemic to defeat. Convincing individuals that a vaccine designed in a specific country (say the US or Russia, for instance) will feed conspiracy theories and other narratives, including its potential lack of testing and fast-track design. A vaccine for COVID-19 would raise many questions, and people often find miss-guiding, incomplete or simply wrong answers to all of them on social media. Anti-vaccination narratives will offer a wide range of attractive and seductive answers which could pull undecided individuals, with genuine questions and concerns about a vaccine, into their anti-vaccination views.

Strictly enforced vaccination would blend perfectly with many of the conspiracy theories created around COVID-19 and would create massive chaos, if we take into account the reactions against a non-invasive element, such as a face mask. If a vaccine is available and there is a possibility of offering universal vaccination in a country, chances are that still many individuals will opt to refuse it. In turn, the time needed to reach herd immunity might increase substantially. If a vaccine becomes available, clear and concise evidence-based communication to the wide audience will be a crucial factor.

Although pandemic-denying, opposing quarantines or face masks and anti-vaccine views are substantially different and are not necessarily shared by the same people, a pandemic-denier, for instance, will most likely oppose being vaccinated against COVID-19, and likely has as many contacts as she or he had before the pandemic, without keeping some physical distance with others or other safety measures (such as a face mask) and thus, would still pass the virus (to as many contacts) if they get infected. But this is one limitation of our study is precisely assuming that the network in which individuals share anti-vaccination views is the same network in which they pass the virus, which is not true if we consider online communities and online contacts. Still under this assumption, one of the severe challenges with anti-vaccine views is that they might be shared by individuals with high centrality in the social network, as they are more exposed to them. Our results suggest that individuals with higher centrality are more likely to share anti-vaccination views, which is detrimental for the results of an optimal vaccination strategy. Central nodes play a highly relevant role both on the diffusion of anti-vaccination views and the evolution of the pandemic. 

Thus, it is worth considering an efficient communication campaign as having a similar impact as a vaccination strategy among the most central nodes. Social media celebrities, Instagram influencers or YouTube stars could help promote evidence-based views about the vaccine but could also act as hubs of miss-information. An effective communication strategy, targeting central nodes with shreds of evidence about vaccines in general, and about COVID-19, could be viewed as an immunisation strategy against fake news which could percolate the network.

\bibliographystyle{unsrt}

\end{document}